\documentclass[onecolumn,prl, preprint, superscriptaddress]{revtex4-1}
\usepackage{bm}
\usepackage[colorlinks=true,linkcolor=blue,citecolor=blue]{hyperref}
\usepackage{times}
\usepackage{amsmath}
\usepackage{amssymb}
\usepackage{amsthm}
\usepackage{amsfonts}
\usepackage{enumerate}
\usepackage{latexsym}
\usepackage{ifpdf}
\newcommand{\beq}{\begin{equation}}
\newcommand{\eeq}{\end{equation}}
\usepackage{graphicx}
\usepackage{makeidx}
\usepackage{color}

\usepackage[version=4]{mhchem}
\usepackage{textcomp,mathcomp}

\begin{document}

\title{Local structural distortions drive magnetic molecular field in compositionally complex spinel oxide}
\author {Rukma Nevgi}
\altaffiliation{Contributed equally}
\email{rukmagurudas@iisc.ac.in}
\affiliation  {Department of Physics, Indian Institute of Science, Bengaluru  560012, India}
\author {Subha Dey}
\altaffiliation{Contributed equally}
\affiliation  {Department of Physics, Indian Institute of Science, Bengaluru 560012, India}
\author {Nandana Bhattacharya}
\affiliation{Department of Physics, Indian Institute of Science, Bengaluru 560012, India}
\author{Soheil Ershadrad}
\affiliation {Department of Physics and Astronomy, Uppsala University, Box-516, 75120 Uppsala, Sweden}
\author{Tinku Dan}
\affiliation  {Deutsches Elektronen-Synchrotron DESY, 22607 Hamburg, Germany}
\author{Sujay Chakravarty}
\affiliation {UGC-DAE CSR Kalpakkam Node Kokilamedu-603104, Tamilnadu, India}
\author{S. D. Kaushik}
\affiliation {UGC-DAE Consortium for Scientific Research Mumbai Centre, R5 Shed, Bhabha Atomic Research Centre, Mumbai 400085, India}
\author{Christoph Klewe}
\affiliation  {Advanced Light Source, Lawrence Berkeley National Laboratory, Berkeley, California 94720, USA}
\author{George E. Sterbinsky}
\affiliation {Advanced Photon Source, Argonne National Laboratory, Lemont, Illinois 60439, USA}
\author{Biplab Sanyal}
\affiliation {Department of Physics and Astronomy, Uppsala University, Box-516, 75120 Uppsala, Sweden}
\author {Srimanta Middey}
\email{smiddey@iisc.ac.in}
\affiliation {Department of Physics, Indian Institute of Science, Bengaluru 560012, India}

\begin{abstract}
{Understanding how local distortions determine the functional properties of high entropy materials, containing  five or more elements at the same crystallographic site, is an open challenge. We address this for a compositionally complex spinel oxide (Mn$_{0.2}$Co$_{0.2}$Ni$_{0.2}$Cu$_{0.2}$Zn$_{0.2}$)Cr$_2$O$_4$ ($A^5$Cr$_2$O$_4$). By comparatively examining extended X-ray absorption fine structure on $A^5$Cr$_2$O$_4$  and its parent counterparts $A$Cr$_2$O$_4$ along with density functional theory calculations for multiple configurations,
we find that the element-specific distortions go beyond the first neighbor. Specifically, the strong Jahn-Teller distortion present in CuCr$_2$O$_4$ is found to be completely suppressed in  $A^5$Cr$_2$O$_4$ even locally. Instead, there is a broad distribution of Cu-O and Cu-Cr bond distances while other $A$-O distances acquire certain specific values.
This study demonstrates the additional flexibility of a cationic sublattice in maintaining a uniform long-range structure, in contrast to previous reports showing only the accommodative anionic sublattice. Remarkably, despite the presence of multiple magnetic ions and variable bond lengths, the mean-field magnetic interactions of $A^5$Cr$_2$O$_4$ exhibit a striking resemblance to those of NiCr$_2$O$_4$. This compelling observation originates from the comparability of bond lengths around Cr in both materials. Our study paves the way for a deeper understanding of the impact of local structural distortions in compositionally complex quantum materials, enabling the targeted design with tailored properties. }

\end{abstract}

\maketitle

 \pagebreak

The periodic arrangement of atoms/ions/molecules within crystalline materials is the backbone of their diverse electronic, magnetic, and topological properties. Conventionally, crystalline materials are synthesized based on the principle of enthalpy minimization. However, in recent years, there has been a surge of interest in high-entropy materials that break this paradigm, where high configurational entropy drives structure formation~\cite{George:2019p515,Sarkar:2019p1806236,Oses:2020p295,Musico:2020p040912,Brahlek:2022p110902,Kotsonis:2023p5587,Mazza:2024embracing}. Originally introduced for multicomponent alloys~\cite{Cantor:2004p213,Yeh:2004p299}, this concept was extended to oxide ceramics in 2015 by demonstrating the stabilization of Mg$_{0.2}$Co$_{0.2}$Ni$_{0.2}$Cu$_{0.2}$Zn$_{0.2}$O in a rock-salt structure~\cite{Rost:2015p8485}. Since then, high entropy oxides (HEOs) have been synthesized with a variety of structures, such as fluorites ($A$O$_{2-\delta}$)~\cite{Djenadic:2017p102}, perovskites ($AB$O$_3$)~\cite{Jiang:2018p116,Sharma:2018p060404,Patel:2020p071601,Brahlek:2020p054407}, spinels ($AB_2$O$_4$)~\cite{Musico:2019p104416,Sharma:2021p17971,Johnstone:2022p20590}, and pyrocholores ($A_2B_2$O$_7$)~\cite{Wright:2020p76,Jiang:2020p4196,Kinsler:2020p104411}, where at least one of the crystallographic sites is occupied by five or more elements in equal or nearly equal atomic fractions. Despite not all being entropy-stabilized~\cite{Brahlek:2022p110902}, these compositionally complex oxides (CCOs) exhibit interesting properties beyond their single element counterparts and hold potential for applications such as energy storage, catalysis, and microwave absorption, etc.~\cite{Sarkar:2018p3400,Chen:2018p11129,Braun:2018p1805004,Kante:2023p5329,Patel:2023p031407,Zhao:2023p2210243,Schweidler:2024p1}.

\begin{figure*}
\begin{center}
\includegraphics[width=0.9\columnwidth]{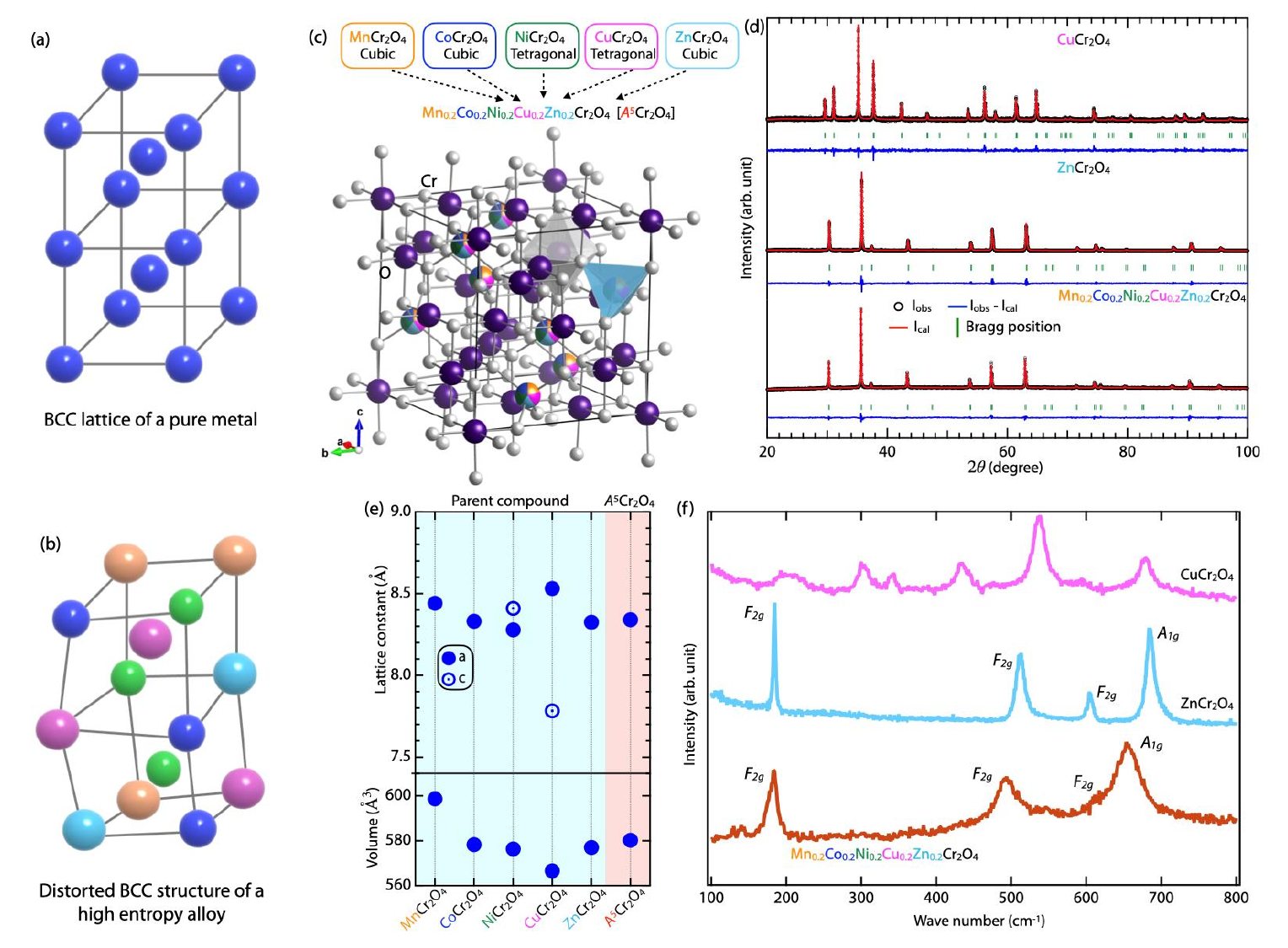}
\caption{\textbf{Long-range structural characterization.}
Schematic to show \textbf(a) body centered cubic (BCC) lattice structure of a pure metal, and \textbf(b) a distorted BCC structure for a high entropy alloy with five five different elements present in near equal fraction.
\textbf{c} Schematic of $A^5$Cr$_2$O$_4$ with cubic spinel structure having Cr at the octahedral site and $A$ (= Mn, Co, Ni, Cu and Zn) at the tetrahedral site.
\textbf{d} Rietveld refined XRD data for CuCr$_2$O$_4$ and ZnCr$_2$O$_4$ along with $A^5$Cr$_2$O$_4$. \textbf{e} The lattice parameters for the parent compounds and $A^5$Cr$_2$O$_4$ as obtained from the Rietveld analysis. The parameter $a$ in the tetragonal phases of NiCr$_2$O$_4$ and CuCr$_2$O$_4$ have been multiplied by $\sqrt{2}$ for ease of comparison. \textbf{f} Raman spectra of $A^5$Cr$_2$O$_4$ has been compared with cubic ZnCr$_2$O$_4$ and tetragonal CuCr$_2$O$_4$. The allowed Raman modes for cubic and tetrogonal spinels are $A_{1g}$+$E_g$+3$F_{2g}$, and 2$A_{1g}$+$B_{1g}$+3$B_{2g}$+4$E_g$, respectively~\cite{DIppolito:2015p1255,Takubo:2011p094406}. In our experiments, we found 4 modes for cubic and 8 modes for tetragonal compounds.}
\label{Fig1}
\end{center}
\end{figure*}

 Local lattice distortions [Fig. ~\ref{Fig1}(a),(b)] in high entropy alloys (HEAs),  a critical factor influencing their mechanical and physical properties,  remain an open issue~\cite{Moniri:2023p564}. The presence of multiple sublattices adds further challenges  in comprehending the interplay among disorder, distortions, and electronic/magnetic properties in CCO ~\cite{Aamlid:2023p5991}. In conventional oxides with low disorder, distortions around individual cations are generally related to their ionic radii, coordination numbers, and oxidation states, which control electron hopping and magnetic exchange interaction strength. The local distortions around each cation within the disordered sublattice of CCO are expected to be highly variable and likely to deviate significantly from the average long-range structure probed by the diffraction technique. It is also necessary to determine the extent of this local distortion variation across neighboring elements and how it affects the material's properties. To address these issues, we employ the extended X-ray absorption fine structure (EXAFS) technique, in conjunction with density functional theory (DFT) calculations.
 EXAFS is an element-specific method used to study the local chemical distribution (length scale up to $\sim$ 5-6 \AA) around particular atoms~\cite{Ruffoni:2007p421, Newville:2014p33,Rehr:2000p621}. EXAFS spectra are specifically sensitive to the coordination number, bond distances, and atomic species surrounding the absorber, providing a relatively simple way to understand the local structural distortion. Interestingly, EXAFS studies have shown that the variation in cation-specific bond lengths is confined only to the first coordination shell in the case of rock-salt HEOs~\cite{Rost:2017p2732,Pu:2023peadi8809}.
 This work rigorously investigates the crucial influence of local structural distortions on the magnetic behavior of a spinel oxide within a compositionally complex setting.

The normal $AB_2$O$_4$ spinel consists of a diamond lattice formed by the tetrahedrally coordinated $A$ site and a pyrochlore lattice formed by the octahedrally coordinated $B$ site [Fig. ~\ref{Fig1}(c)].
The spinel family has been extensively studied for over a century due to the observation of various phenomena, such as magnetic ordering, frustrated magnetism, orbital ordering, charge ordering, metal-insulator transitions, etc~\cite{Tsurkan:2021p1}. This study centers on the $A$Cr$_2$O$_4$ family, a group of normal spinel compounds that display diverse low-temperature magnetic behaviors.   Interestingly, their Curie-Weiss temperatures ($\theta_\mathrm{CW}$), ranging from -400 K to 200 K, directly correlate with the Cr-Cr separation~\cite{Rudolf:2007p76}.
In this work, we have synthesized (Mn$_{0.2}$Co$_{0.2}$Ni$_{0.2}$Cu$_{0.2}$Zn$_{0.2})$Cr$_{2}$O$_{4}$ (hereafter referred to as $A^5$Cr$_2$O$_4$). The individual members MnCr$_2$O$_4$ and CoCr$_2$O$_4$ are cubic and become ferrimagnetic at $T_c$ values of 41 K and 93 K, respectively~\cite{Tomiyasu:2004p214434,Bhowmik:2006p144413,Yamasaki:2006p207204, Dey:2014p184424}. Additionally, they exhibit multiferroic behavior below 18 K and 27 K~\cite{Tomiyasu:2004p214434,Bhowmik:2006p144413,Yamasaki:2006p207204, Dey:2014p184424}. Due to the Jahn-Teller activity of the Ni$^{2+}$ and Cu$^{2+}$ ions, NiCr$_2$O$_4$ and CuCr$_2$O$_4$ undergo cubic to tetragonal transitions below 310 K and 853 K, respectively. They transform into an orthorhombic phase at 65 K and 125 K, respectively, accompanied by a ferrimagnetic transition~\cite{Suchomel:2012p054406}. On the other hand, ZnCr$_2$O$_4$ is cubic at room temperature and exhibits an antiferromagnetic ordering transition below 12 K~\cite{Dutton:2011p064407}.

In the present work, we probe the modification of the local structure of $A^5$Cr$_2$O$_4$ compared to that of its five undoped counterparts ($A$Cr$_2$O$_4$, $A$ = Mn, Co, Ni, Cu, Zn) at room temperature. X-ray diffraction (XRD), Raman spectroscopy, and X-ray absorption spectroscopy (XAS) of the transition metal $L_{3,2}$ edges confirmed the desired normal spinel structure of $A^5$Cr$_2$O$_4$ with cubic symmetry. Our comprehensive EXAFS measurements and DFT based calculations revealed the element-specific local structural distortions in $A^5$Cr$_2$O$_4$ persist beyond the first neighbor.
Interestingly, the Jahn-Teller distortions typically observed in the CuCr$_2$O$_4$ are absent in $A^5$Cr$_2$O$_4$. Rather, CuO$_4$ units are highly flexible to accommodate local distortions while maintaining a uniform long-range cubic structure. We further show that the near-identical Cr-O and Cr-Cr bond lengths in $A^5$Cr$_2$O$_4$ and NiCr$_2$O$_4$ are responsible for their similar $\theta_\mathrm{CW}$.

\begin{figure}
\begin{center}
\includegraphics[width=1\columnwidth]{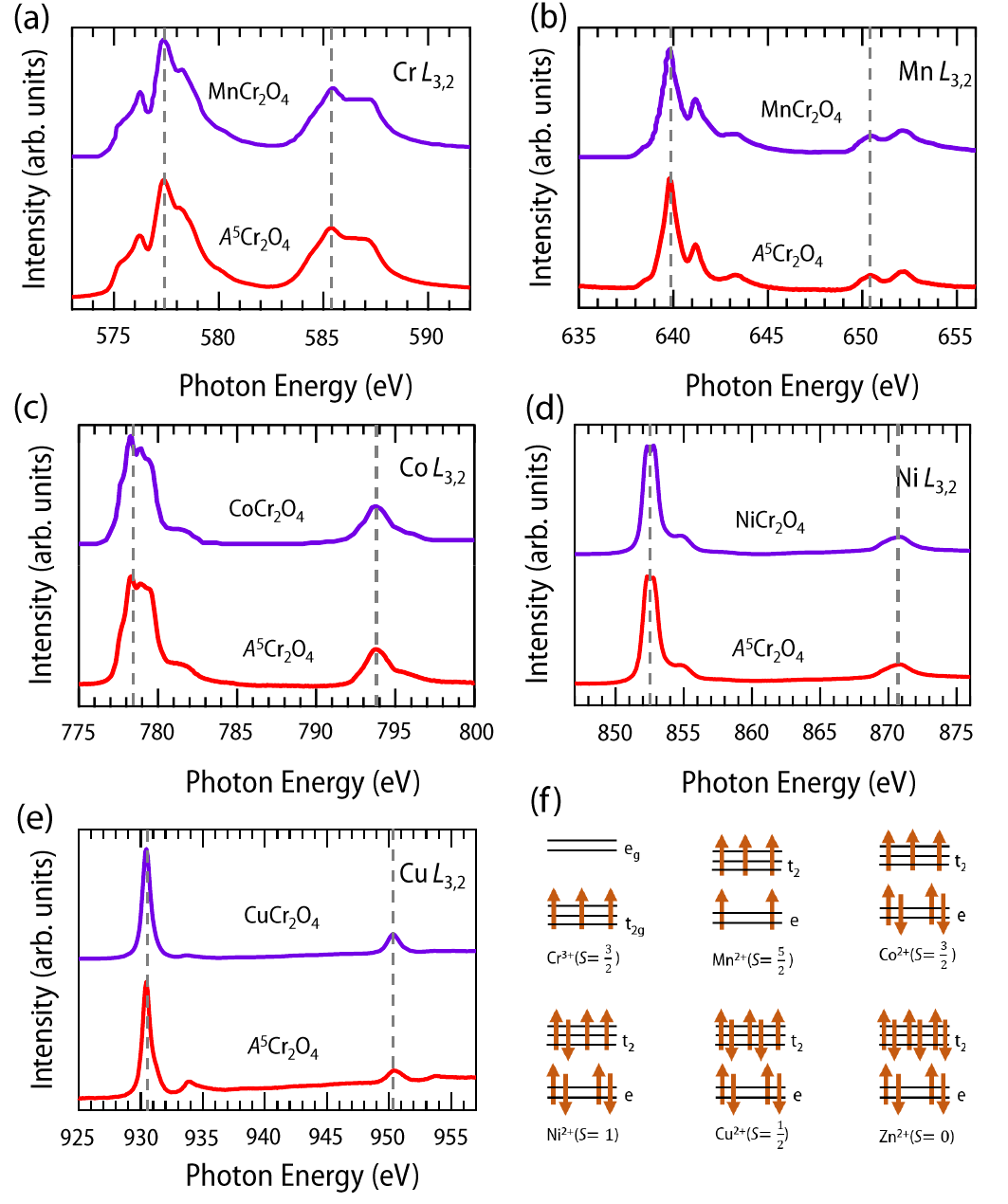}
\caption{\textbf{Determination of oxidation state using X-ray absorption spectroscopy.} \textbf{a} Cr $L_{3,2}$ edge in MnCr$_2$O$_4$ and $A^5$Cr$_2$O$_4$. \textbf{b} Mn $L_{3,2}$ edge in MnCr$_2$O$_4$ and $A^5$Cr$_2$O$_4$. \textbf{c} Co $L_{3,2}$ edge in CoCr$_2$O$_4$ and $A^5$Cr$_2$O$_4$. \textbf{d} Ni $L_{3,2}$ edge in NiCr$_2$O$_4$ and $A^5$Cr$_2$O$_4$. \textbf{e} Cu $L_{3,2}$ edge in CuCr$_2$O$_4$ and $A^5$Cr$_2$O$_4$. The spectra of Cr and Mn for MnCr$_2$O$_4$ is taken from Ref. ~\onlinecite{Vanderlaan:2010p067405}. The spectra of Co for CoCr$_2$O$_4$ is adapted from Ref.~\onlinecite{kim:2009p94}. \textbf{f} Spin configurations of Cr, Mn, Co, Ni, Cu, and Zn cations in $A^5$Cr$_2$O$_4$. The energy gap between $t_{2g}$ ($t_2$) and $e_g$ ($e$) is not according to the scale.}
\label{fig: Fig 2}
\end{center}
\end{figure}

\noindent \textbf{Results}:

{\color{magenta}\bf Synthesis and global structural symmetry:}
The synthesis of polycrystalline samples of five parent compounds $A$Cr$_2$O$_4$  (\textit{A} = Mn, Co, Ni, Cu and Zn) and  $A^5$Cr$_2$O$_4$ was carried out using a conventional solid-state synthesis route (details are provided in the Methods section). The powder X-ray diffraction (XRD) pattern of $A^5$Cr$_2$O$_4$ at room temperature is compared with that of two parent members, tetragonal CuCr$_2$O$_4$ and cubic ZnCr$_2$O$_4$, in Fig.~\ref{Fig1}(d). Similar to ZnCr$_2$O$_4$, the diffraction pattern of $A^5$Cr$_2$O$_4$ can be indexed and refined with a normal spinel structure with cubic symmetry having a space group of $Fd\bar{3}m$ (space group number: 227).
The diffraction patterns and structural information obtained from the refinement of all six compounds are shown in the Supplementary (Table S1 and Fig. S1).
The lattice constants and lattice volumes of five parent members and  $A^5$Cr$_2$O$_4$ at room temperature are shown in Fig.~\ref{Fig1}(e). For comparison, the lattice constant $a$ of the tetragonal members NiCr$_2$O$_4$ and CuCr$_2$O$_4$ is multiplied by $\sqrt{2}$ (see Supplementary Fig. S2). The trend in lattice parameters and volume can be understood by considering the ionic radii of $A^{2+}$ ions. The average ionic radii (0.592 \AA) of the $A$-site for $A^5$Cr$_2$O$_4$ are closer to those of Zn$^{2+}$ (0.60 \AA) and Co$^{2+}$ (0.58 \AA) [ionic radii of Mn$^{2+}$: 0.66 \AA, Ni$^{2+}$: 0.55 \AA, and Cu$^{2+}$: 0.57 \AA]~\cite{Shannon:1970p1046}, leading to similar unit cell volumes of $A^5$Cr$_2$O$_4$ and $A$Cr$_2$O$_4$ ($A$= Co and Zn). Although NiCr$_2$O$_4$ is tetragonal, the Jahn-Teller distortion is very small at 300 K, and the unit cell volume is very similar to $A^5$Cr$_2$O$_4$. It can also be inferred from Fig.~\ref{Fig1}(e) that the Mn and Cu ions are likely to experience internal stress (compressive for Mn and tensile for Cu) when introduced into $A^5$Cr$_2$O$_4$, leading to local structural distortions.

The Raman spectrum of $A^5$Cr$_2$O$_4$ has been further compared with cubic ZnCr$_2$O$_4$ and tetragonal CuCr$_2$O$_4$ at Fig.~\ref{Fig1}(f). ZnCr$_2$O$_4$ shows $F_{2g}$(1), $F_{2g}$(2), $F_{2g}$(3) and $A_{1g}$ modes at 185 cm$^{-1}$, 511 cm$^{-1}$, 604 cm$^{-1}$, 684 cm$^{-1}$, respectively, similar to earlier report~\cite{DIppolito:2015p1255}. Due to the lower symmetry, CuCr$_2$O$_4$ exhibits additional Raman modes~\cite{Takubo:2011p094406}.  The similarity between the Raman spectra of $A^5$Cr$_2$O$_4$ and ZnCr$_2$O$_4$ further establishes its cubic symmetry. More significantly, we observe shifts and broadening of Raman modes, providing direct evidence of substantial lattice distortions within the compositionally complex $A^5$Cr$_2$O$_4$~\cite{Krysko:2023p101123}, which we probe further in detail using element-sensitive EXAFS technique.

\begin{figure}
\begin{center}
\includegraphics[width=.95\columnwidth]{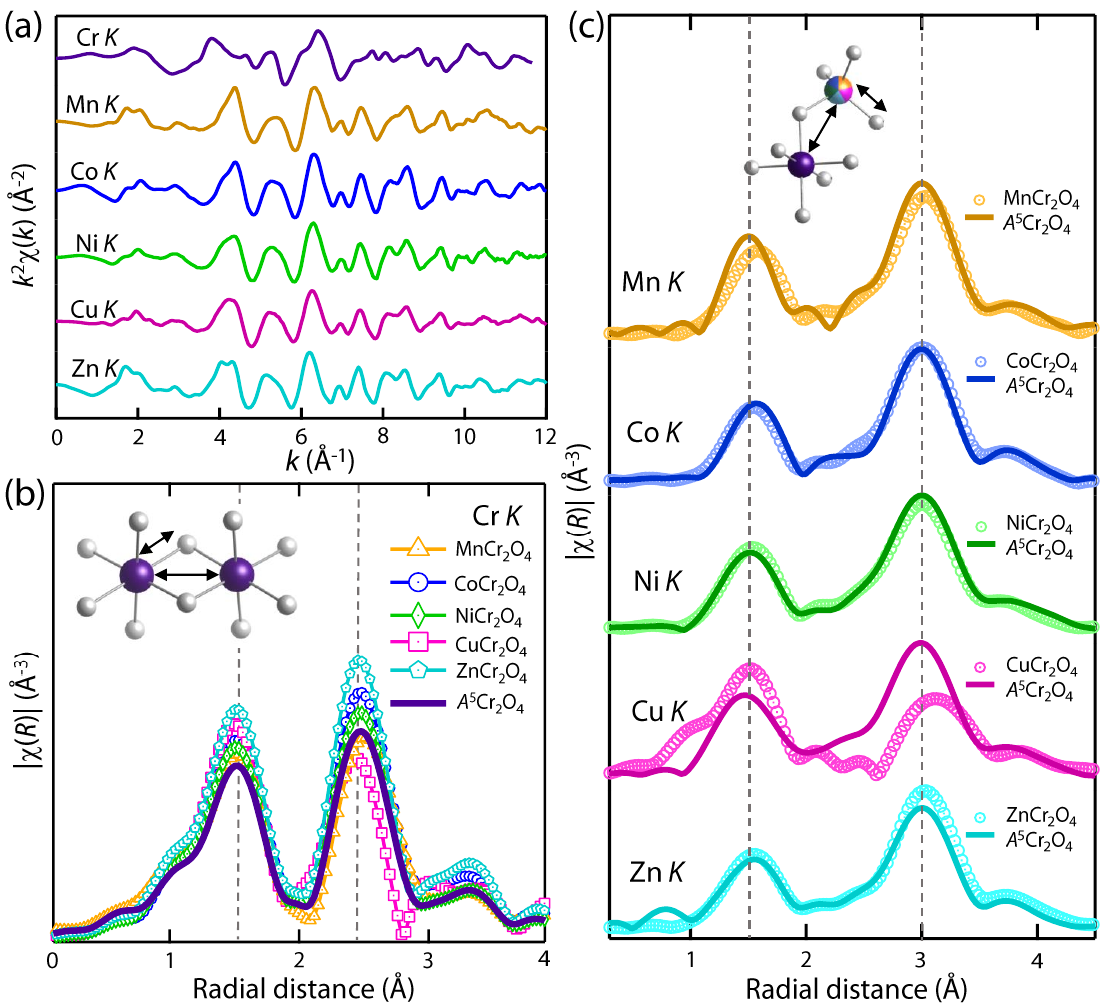}
\caption{\textbf{Probing the local structure using EXAFS.} \textbf{a} The \textit{k} weighted EXAFS spectra in $A^5$Cr$_2$O$_4$. The Fourier transform of the EXAFS spectra at \textbf{b} Cr $K$-edge and \textbf{c} $A$ $K$-edges in the parent oxides and $A^5$Cr$_2$O$_4$. Inset panels show the first and second nearest neighbor distances (b) for Cr, (c) for $A$ site.}
\label{fig: Fig 3}
\end{center}
\end{figure}

{\color{magenta}\bf Transition metal valency and crystal field environment:}
Prior to investigating local distortions, we determine the location of cations within octahedral and tetrahedral sites. For this, we carried out element-specific XAS experiments on the $L_{3,2}$ edges of transition metal ions because the XAS spectral line shape for the 2$p^6$3$d^n\rightarrow$2$p^5$3$d^{n+1}$ transition is strongly dependent on the valency, spin character of the initial state and crystal field environment of the system~\cite{Stohr:2006book}.
The XAS spectra for the $L_{3,2}$ edges of Cr, Mn, Co, Ni, and Cu in $A^5$Cr$_2$O$_4$, recorded in total electron yield mode, were compared with those of the corresponding parent compounds, as shown in Fig.~\ref{fig: Fig 2} (a)-(e). In all parent $A$Cr$_2$O$_4$ compounds, the cations $A^{2+}$ occupy the tetrahedral site, while Cr$^{3+}$ occupies the octahedral site~\cite{Tsurkan:2021p1}. It is evident from Fig.~\ref{fig: Fig 2} (a)-(e) that the spectra of Cr, Mn, Co, and Ni for $A^5$Cr$_2$O$_4$ look exactly similar to the corresponding edges of the parent compounds~\cite{Windsor:2017p224413, Vanderlaan:2010p067405, kim:2009p94}.
Although the main features of Cu $L_{3,2}$ XAS for $A^5$Cr$_2$O$_4$ are quite similar to those of the parent CuCr$_2$O$_4$, the features at approximately 933.9 eV and 953.7 eV are more intense. This slight difference may be attributed to the differences in crystal field parameters due to the variation in local structures around Cu (demonstrated in the latter part of the manuscript). Overall, our XAS measurements confirmed that the compound has a normal spinel structure, similar to that of its parent counterparts. Fig.~\ref{fig: Fig 2}(f) shows the expected spin configurations for each of the cations of $A^5$Cr$_2$O$_4$, according to the oxidation state found by XAS. This is further corroborated by the Curie‒Weiss analysis of magnetic susceptibility, which is discussed in the later section of the manuscript.

{\color{magenta}\bf Investigation of local distortions:} To investigate the local structural distortions around each cation in $A^5$Cr$_2$O$_4$, we analyzed EXAFS data for each of the metal \textit{K}-edge in $A^5$Cr$_2$O$_4$ as well as all the parent compounds (the details of experiment and data analysis have been provided in Method section). 
 Supplementary Fig. S3 displays XANES (X-ray absorption near edge structure) region of the $K$-edge XAS spectra, confirming octahedral coordination of Cr and tetrahedral coordination of $A$ elements. The \textit{K} edge EXAFS spectra are shown in Fig.~\ref{fig: Fig 3}.  As anticipated, the Mn, Co, Ni, Cu, and Zn $K$-edge EXAFS features [Fig.~\ref{fig: Fig 3}(a)] are similar, but distinct from the Cr $K$-edge.

The Fourier transform (FT) of $k^2.\chi(k)$ of Cr and $A$ $K$-edges for each of the parent compounds and $A^5$Cr$_2$O$_4$ are shown in Fig.~\ref{fig: Fig 3}(b) and (c), respectively. The first peak at 1.55 \AA \ in both Cr and $A$ $K$-edge data suggests similar Cr-O and $A$-O distances. The second peak at 2.5 \AA \ in Cr $K$-edge data corresponds to Cr-Cr distance, while the one at 3 \AA \ in $A$ $K$-edge data corresponds to $A$-Cr distance, implying different cation environments around Cr and the \textit{A} cations. We also note that the real scattering distance shifts by approximately 0.5 \AA\, as the phase shift in the Fourier transform remains uncorrected.

The FT of $k^2.\chi(k)$ for Cr $K$ in $A^5$Cr$_2$O$_4$ [Fig.~\ref{fig: Fig 3}(b)] shows similar features like the parent compounds (except CuCr$_2$O$_4$). However, Fig.~\ref{fig: Fig 3}(c) reveals significant changes in local structures around Cu and Mn in $A^5$Cr$_2$O$_4$ compared to CuCr$_2$O$_4$ and MnCr$_2$O$_4$, respectively. This is likely due to the difference in crystal symmetry (cubic vs. tetragonal) and lattice volume changes. Local structures around Co, Ni, and Zn in $A^5$Cr$_2$O$_4$ are similar to their parent compounds, except for a slight change in Zn-Cr distance. To quantify all differences in local structures,
we performed fittings for the Cr \textit{K} edge and $A$ $K$-edge ($A$ = Mn, Co, Ni, Cu and Zn) spectra, as discussed below.

\begin{figure}
\begin{center}
\includegraphics[width=0.8\textwidth]{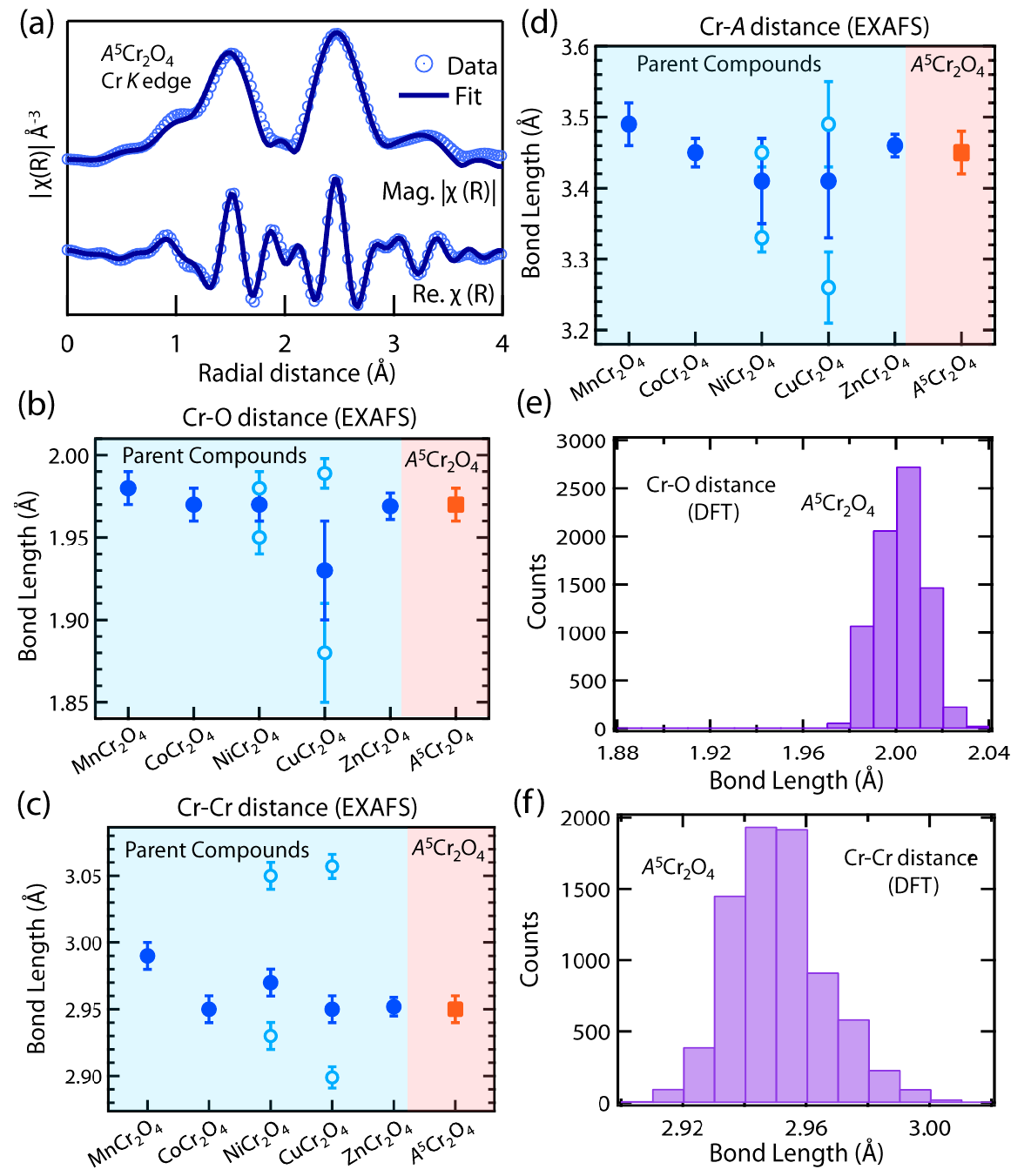}
\caption{\textbf{Analysis of local structure in the octahedral environment.}
\textbf{a} The Fourier transformed EXAFS spectrum along with the fitted data [the magnitude $|\chi (R)|$ and the real part Re. $\chi (R)$]  in $A^5$Cr$_2$O$_4$. The bond distances from Cr for \textbf{b} first neighbor O, \textbf{c} second neighbor Cr and \textbf{d} third neighbor $A$ ions as obtained from EXAFS fittings for every pristine chromite spinels and $A^5$Cr$_2$O$_4$. Due to Jahn-Teller distortion, NiCr$_2$O$_4$ and CuCr$_2$O$_4$ display two bond lengths (drawn as open circles),  whose weighted average  (denoted by the closed circle) is also plotted for ease of comparison.  Histogram plots for \textbf{e} Cr-O,  and \textbf{f} Cr-Cr bond lengths, obtained from DFT calculations of 10 different disordered structures.}
\label{fig: Fig 4}
\end{center}
\end{figure}

\begin{figure*}
\begin{center}
\includegraphics[width=0.8\textwidth ]{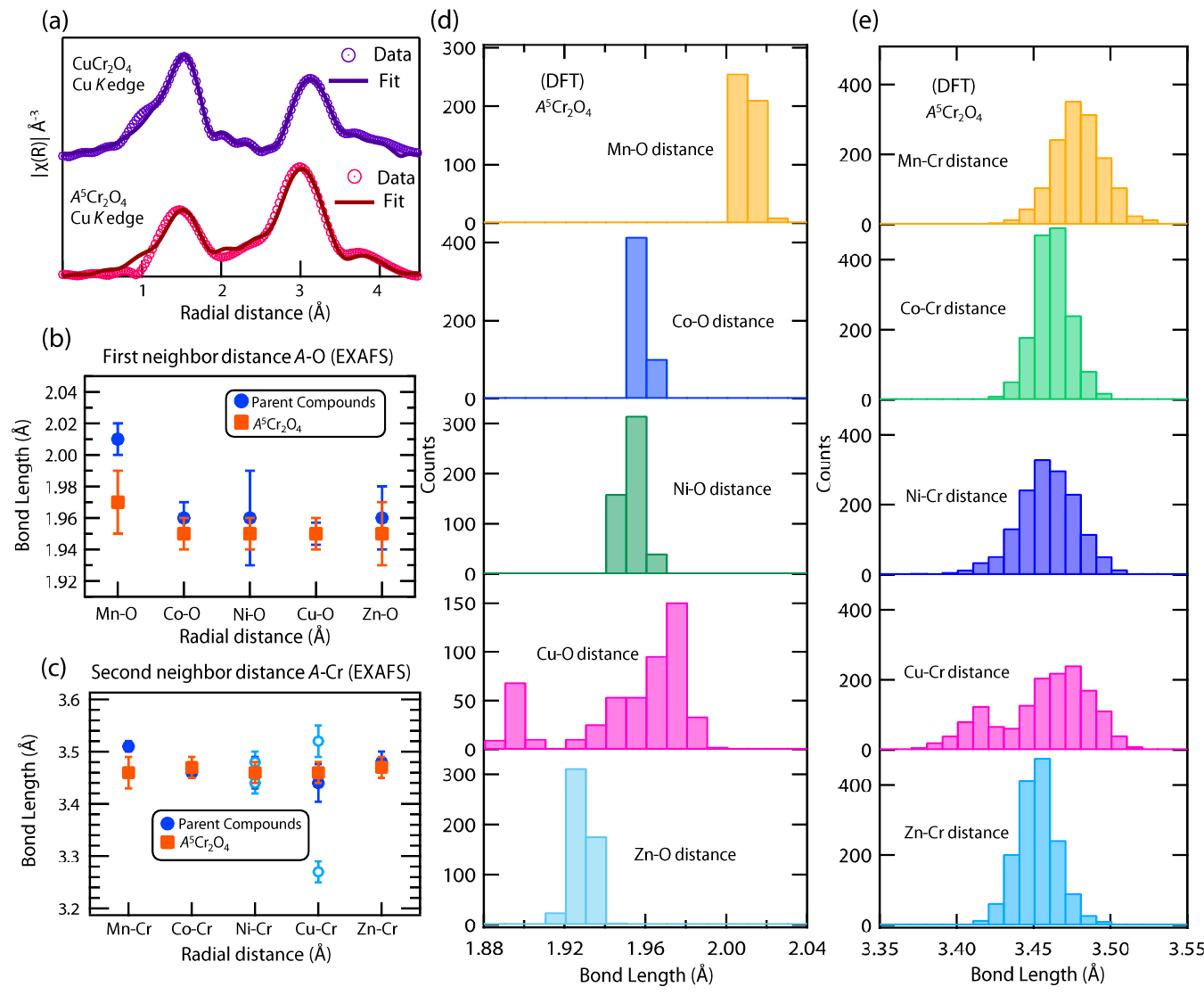}
\caption{\textbf{Analysis of local structure in the tetrahedral environment.}The magnitude of the Fourier transformed EXAFS spectra along with fittings obtained at the \textbf{a} Cu $K$-edge in CuCr$_2$O$_4$ and $A^5$Cr$_2$O$_4$. The bond lengths depicted in [\textbf{b}-\textbf{c}] are obtained from EXAFS fittings for every pristine chromite spinels and $A^5$Cr$_2$O$_4$. In \textbf{c}, due to Jahn-Teller distortion, NiCr$_2$O$_4$, and CuCr$_2$O$_4$ display two bond lengths between $A$ and Cr cations (drawn as open circles), whose weighted average (closed circle) is plotted for ease of comparison.  Histogram plots for \textbf{d} first neighbor $A$-O,  and \textbf{e} second neighbor $A$-Cr bond lengths, found from DFT calculations of 10 different disordered structures.}
\label{fig: Fig 5}
\end{center}
\end{figure*}

{\bf Local structure around Cr:}
For the Cr \textit{K}-edge EXAFS, we employed a \textit{k} range of 2 to 12 \AA$^{-1}$ and an \textit{R} range of 1 to 3.7 \AA\ for the fittings (see Fig. 4(a), Fig. S4, and Tables S2-S4 in the Supplementary).
All analyses were performed considering the structure symmetries determined by XRD: tetragonal for CuCr$_2$O$_4$ and NiCr$_2$O$_4$, and cubic for the remaining  parent compounds and $A^5$Cr$_2$O$_4$. In the cubic model, the Cr site has 6 O as the first neighbor, 6 Cr as the second neighbor, and 6 $A$ cations as the third neighbor.
The bond lengths obtained from the EXAFS fittings have been shown in Fig.~\ref{fig: Fig 4}(b)-(d), respectively.
In the tetragonal structure, the six equivalent bond lengths split into two distinct groups: four with one distance and two with another. Therefore, for NiCr$_2$O$_4$ and CuCr$_2$O$_4$, the weighted average (denoted by closed circles) is also plotted for easier comparison in Fig.~\ref{fig: Fig 4}(b)-(d). From the analysis, we could infer that the bond distances Cr-O (1.97 \AA), Cr-Cr (2.96 \AA) and Cr-$A$ (3.46 \AA) in $A^5$Cr$_2$O$_4$ are akin to that of CoCr$_2$O$_4$, ZnCr$_2$O$_4$ and also with the weighted average bond lengths of NiCr$_2$O$_4$. Consistent with the difference in lattice constant, these bond lengths are higher in MnCr$_2$O$_4$. In addition, the average Cr-O and Cr-$A$ bond distances of CuCr$_2$O$_4$ are lower than $A^5$Cr$_2$O$_4$. Since CuCr$_2$O$_4$ and NiCr$_2$O$_4$ exhibit tetragonal structures at 300 K, we attempted to analyze the Cr K-edge EXAFS of $A^5$Cr$_2$O$_4$ using tetragonal symmetry. However, this yielded unphysical fitting parameters (Table S6, Supplementary), signifying the absence of local Jahn-Teller distortions around Cr.

  To further examine any local Jahn-Teller like distortions around Cr, we performed DFT calculations on 10 supercells, each containing 448 atoms. These supercells were generated by randomly distributing 12 Mn, 13 Co, 13 Ni, 13 Cu, and 13 Zn atoms across the 64 $A$-sites, creating 10 distinct disordered structures (calculation details are in the Methods section). Histograms of Cr-O and Cr-Cr bond distances, derived from these calculations, are shown in Fig.~\ref{fig: Fig 4}(e) and (f), respectively. While a narrow distribution of Cr-O bond distances ($\sim$2.0$\pm$0.02 \AA) was observed, no Jahn-Teller distortions were found (our DFT calculations accurately capture Jahn-Teller distortions in CuCr$_2$O$_4$, as shown in Supplementary Table S7). The Cr-Cr bond distances also exhibit a distribution, with the peak position [Fig.~\ref{fig: Fig 4}(f)] closely matching the EXAFS-derived value for $A^5$Cr$_2$O$_4$ [Fig.~\ref{fig: Fig 4}(c)]. However, EXAFS is limited in revealing precise distributional information due to its averaging of local environments around all Cr ions.

{\bf Local structure around $A$ sites:}
To probe the local distortions around $A$-sites,
we performed fittings for the $A$ $K$-edge spectra ($A$ = Mn, Co, Ni, Cu and Zn) in every parent compound and $A^5$Cr$_2$O$_4$ in the $k$ range of 2 to 14 \AA$^{-1}$ and the $R$ range of 1 to 4.5 \AA\ (details are in the Methods section, see Fig.~\ref{fig: Fig 5}(a) in main text, Fig. S4 and Table S2-S4 in Supplementary). Based on XRD results, either cubic or tetragonal symmetry was used for these fittings and all coordination numbers were also kept fixed. Fits with variable coordination numbers (see Supplementary section S2.3 and Table S5) corroborate the reliability of our fixed coordination number approach.

We first focus on the parent compounds. For the cubic MnCr$_2$O$_4$, EXAFS analysis indicates that Mn is coordinated with 4 O atoms at 1.99 \AA, 12 O atoms at 3.32 \AA, 12 Cr atoms at 3.50 \AA, and 4 Mn atoms at 3.64 \AA. Notably, the Mn-O bond distance, identified as the third neighbor by XRD, was found to be shorter than the Mn-Cr bond distance (second neighbor) in EXAFS fitting. We observe this trend across all six compounds, as well as in previous EXAFS studies~\cite{Acharyya:20144232-4241,Galivarapu:201663809-63819}. This deviation in bond distances is likely due to the weak scattering contribution from O as a third neighbor, a consequence of its low atomic number. Therefore, the subsequent sections of this paper will primarily focus on the first-neighbor $A$-O and second-neighbor $A$-Cr distances. It should be noted that the $A$-Cr distances are equivalent to the Cr-$A$ distances, which were independently determined from Cr $K$-edge EXAFS fitting.
For tetragonal CuCr$_2$O$_4$, EXAFS analysis shows the Cu atom coordinated with 4 O at 1.95 \textup{~\AA}  in the first shell, 8 Cr at 3.52 \AA, and 4 Cr at 3.26 \AA\ in the second shell.


The EXAFS fittings of each $A$-site of $A^5$Cr$_2$O$_4$ were performed considering cubic symmetry in accordance with the XRD result. Excellent fitting has been found for each case, including Cu $K$-edge [Fig.~\ref{fig: Fig 5} (a)] and Ni $K$-edge (see Supplementary Fig. S3 (c)). We have also attempted to fit Cu $K$ EXAFS of $A^5$Cr$_2$O$_4$ using tetragonal symmetry, similar to the parent compound CuCr$_2$O$_4$. However, such fitting results in unphysical parameters (see Supplementary section S2.4, Table S6), indicating the absence of tetragonal distortion, even locally.
The comparison of each neighbor distance among the pristine members and doped compound reveals several interesting trends (Fig. \ref{fig: Fig 5} (b) and (d)).  The Mn-O distance in MnCr$_2$O$_4$ is the largest, attributed to its larger unit cell volume compared to other pristine materials. Notably, the first neighbor $A$-O distance of $A^5$Cr$_2$O$_4$ depends on the specific element positioned at the $A$-site. The Co-O, Ni-O, and Cu-O distances are very similar, whereas Mn-O and Zn-O distances differ. Also, all the $A$-O bond distances are smaller than those in the corresponding parent $A$Cr$_2$O$_4$, indicating that these local variations are not simply due to changes in the overall crystal volume. Furthermore,  second neighbor $A$-Cr distances [Fig. \ref{fig: Fig 5}(c)], determined from $A$-site EXAFS in  $A^5$Cr$_2$O$_4$, vary mildly with the specific $A$ cation.

Next, we discuss the bond lengths at the $A$ site in $A^5$Cr$_2$O$_4$ based on DFT calculations. The histogram plots for the first neighbor $A$-O distances and the second neighbor $A$-Cr distances are presented in Fig. \ref{fig: Fig 5} (d) and (e), respectively. Despite considering ten different disordered structures in our calculations, the distances for Mn-O, Co-O, Ni-O, and Zn-O do not show any statistical distribution,  suggesting well-defined bond lengths.
In contrast, the Cu-O bond distances display a broad range of values. Notably, the trends observed for the $A$-O bond distances closely align with our EXAFS results, with Mn-O being the longest and Zn-O the shortest.
 Interestingly, the second neighbor distances for Mn-Cr, Co-Cr, Ni-Cr, and Zn-Cr also show a distribution, as illustrated in Fig. \ref{fig: Fig 5} (e). The peak positions of these distributions are similar to the Cr-$A$ distances obtained from EXAFS, shown in Fig. \ref{fig: Fig 5} (c). Furthermore, the distribution width for Cu-Cr bonds is significantly broader compared to the other $A$-Cr bonds.
This finding corroborates our EXAFS analysis, which revealed a higher $\sigma^2$
 (mean squared variation of bond length) for Cu-O and Cu-Cr distances compared to other $A$-O and $A$-Cr bonds ( see Supplementary Fig. S5).


{\bf Accommodation of local distortions:} Our EXAFS and DFT analyses demonstrate that element-specific structural distortions in $A^5$Cr$_2$O$_4$ extend beyond the first coordination shell of the $A$ sites. Notably, the broad distribution of Cu-O bond distances, alongside specific values for other $A$-O distances, highlights the crucial role of the highly flexible CuO$_4$ tetrahedral units in accommodating these local distortions.  Consequently, this flexibility influences other bond distances: the connection between CuO$_4$ and CrO$_6$ unit via a common oxygen (inset of Fig.~\ref{fig: Fig 3}(c)) leads to a broad Cu-Cr distance distribution, while Cr-O and Cr-Cr distances show narrower distributions without Jahn-Teller distortion. Additionally, the connectivity of each CrO$_6$ octahedron to five $A$O$_4$ units ($A$ = Mn, Co, Ni, Cu, Zn) contributes to the narrower distributions of other Cr-$A$ bond lengths.

 \begin{figure*}
\begin{center}
\includegraphics[width=0.8\textwidth ]{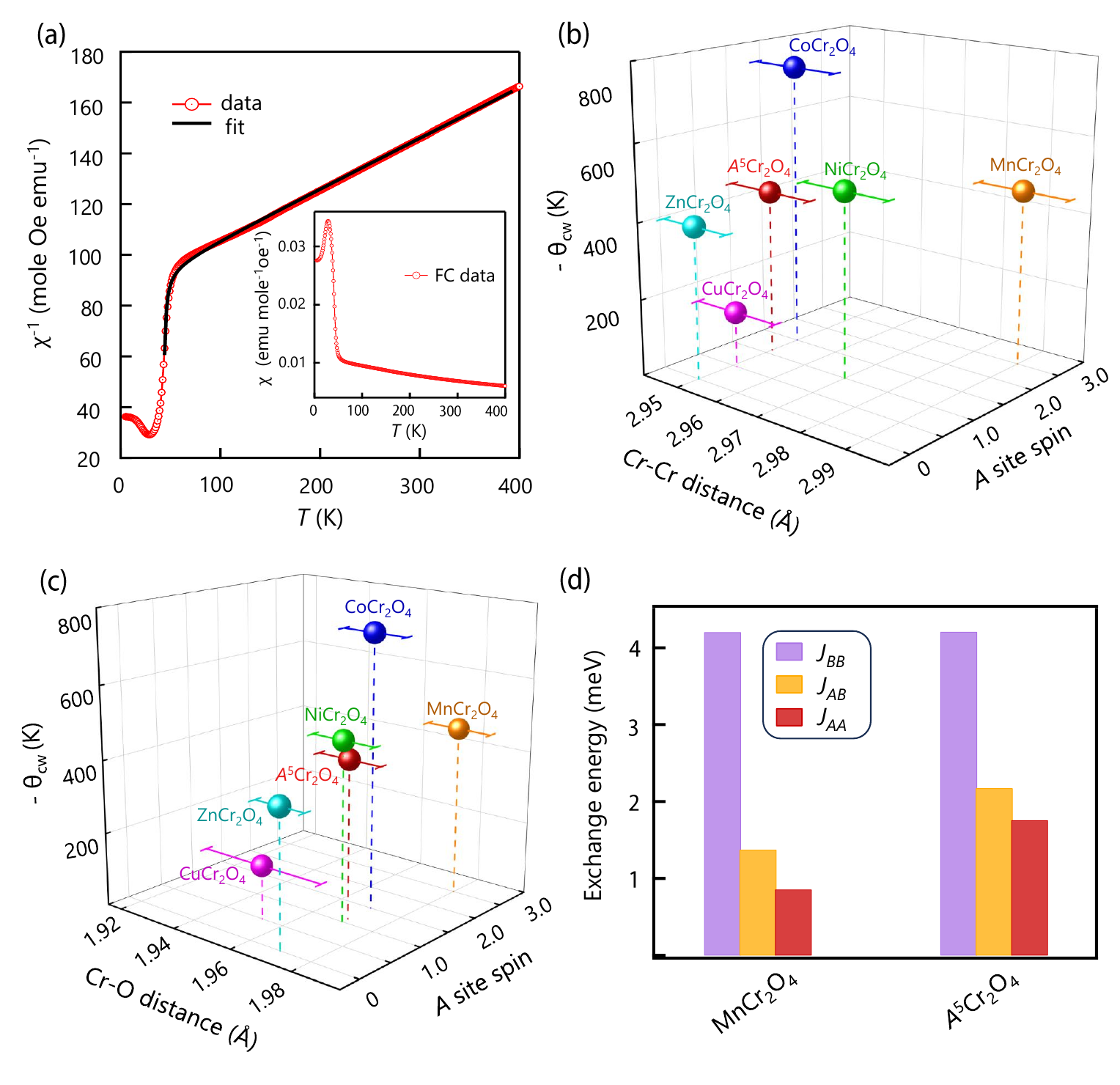}
\caption{\textbf{Magnetic characterization and it's connection with local structure} \textbf{a} Fitting of inverse magnetic dc susceptibility using Curie-Weiss equation above $T_{N}$ in $A^5$Cr$_2$O$_4$ at 5000 Oe. Inset features field cooled (FC) dc susceptibility as a function of temperature. -$\theta_\mathrm{CW}$ is plotted as a function of \textbf{b} Cr-Cr bond distance and $A$-site spin and \textbf{c} Cr-O bond distance and $A$ site spin in parent oxides and $A^5$Cr$_2$O$_4$. \textbf{d} A comparision of the magnetic exchange interaction energy between two Cr cations ($J_{CrCr}$), between Cr and $A$ ($J_{ACr}$), and between two $A$ cations ($J_{AA}$) in MnCr$_2$O$_4$ and $A^5$Cr$_2$O$_4$.}
\label{fig: Fig 6}
\end{center}
\end{figure*}

{\color{magenta}\bf Correlation between structural distortions and magnetic interaction:} These bond lengths are expected to play a crucial role in comprehending the strength of magnetic interactions in $A$Cr$_2$O$_4$ spinels. The theory of ground state magnetic configuration of cubic $AB_2$O$_4$ is very often described by the following Hamiltonian with classical Heisenberg spins considering $B$-$B$, and $A$-$B$ exchanges, initially proposed by Lyons, Kaplan, Dwight, and Menyuk~\cite{Kaplan:2007p3711,Kaplan:1960p1460,Tsurkan:2021p1}.

\begin{equation}
H=2J_{AB}S_AS_B\left(\sum_{\langle ij \rangle}^{A,B}\overrightarrow{\sigma_i^A}.\overrightarrow{\sigma_j^B}+\frac{3u}{4}\sum_{\langle ij \rangle}^{B,B}\overrightarrow{\sigma_i^B}.\overrightarrow{\sigma_j^B}\right)
\end{equation}
The sums are over nearest-neighbor $A$-$B$ and $B$-$B$ pairs and the parameter $u$ =$\frac{4J_{BB}S_B}{3J_{AB}S_{A}}$ and $\overrightarrow{\sigma_j^\alpha}$=$\overrightarrow{S_j^\alpha}$/$S_j^\alpha$ with $\alpha$=$A$ and $B$. The interaction within the diamond sublattice ($J_{AA}$) has also been claimed to be important in recent studies~\cite{Winkler:2009p104418,Ederer:2007p064409}. In chromite spinels, the magnetic interaction between two Cr is a direct exchange and thus depends strongly on the Cr-Cr distance~\cite{Winkler:2009p104418}. The superexchange between $A$ and Cr is mediated through interconnecting oxygen, and the strength will depend on the bond lengths $A$-O and Cr-O,  $A$-O-Cr bond angle, and the electronic configuration of $A$ site ions. Surprisingly, the $\theta_\mathrm{CW}$, which is  a measure of molecular field and considered as an approximate indicator of the strength of mean-field magnetic interaction between the ions~\cite{Mugiraneza:2022p95}, directly correlates with the Cr-Cr distance~\cite{Rudolf:2007p76}. Thus,  we have focused on finding out the relation between $\theta_\mathrm{CW}$ and the bond lengths at 300 K obtained from the EXAFS analysis.

 The temperature-dependent magnetic susceptibility [$\chi$=$M/H$] for $A^5$Cr$_2$O$_4$, measured under a magnetic field ($H$) of 5000 Oe in field cooled condition has been depicted as an inset of Fig. \ref{fig: Fig 6} (a). We found a magnetic transition around 41 K.  The main panel represents the fitting of the inverse of dc susceptibility at high temperature using molecular field theory of a ferrimagnet~\cite{Coey2010}
 \begin{equation}
     \chi^{-1} = \frac{T-\theta_{CW}}{C_{A} + 2C_{B}}-\frac{C^{''}}{T - \theta^{'}}
 \end{equation}
The first term is the hyperbolic high-temperature linear part with a Curie-Weiss form, where $C_A$ and $C_B$ correspond to Curie constants related to the two magnetic sublattices $M_A$ and $M_B$, respectively.
$\theta_\mathrm{CW}$ is the Curie-Weiss temperature. The second term is the hyperbolic low-$T$ asymptote,  where $C^{''}$ and $\theta^{'}$ are constants akin to different Weiss coefficients that represent the inter and intrasublattice interactions.
From this fitting, we have obtained a Curie-Weiss temperature of $\theta_\mathrm{CW}$ = -432 K and the effective magnetic moment of 6.33 $\mu_B$/f.u. using the formula $\mu_\mathrm{eff}$=$\sqrt{\frac{3K_{B}(C_{A}+2C_{B})}{N_{A}}}$~\cite{kassem2023structural}, where $K_{B}$ is Boltzmann's constant and $N_{A}$ is Avogadro's number.
Interestingly, $\mu_\mathrm{eff}$
is very similar to the expected value of 6.26 $\mu_B$/f.u., calculated by considering the average magnetic moments of $A^{2+}$ ($A$ = Mn, Co, Ni, Cu; Zn$^{2+}$ is nonmagnetic) and Cr$^{3+}$  of $A^5$Cr$_2$O$_4$ ($\mu_\mathrm{eff}$=$\sqrt{2\mu_\mathrm{Cr}^2+\mu_{A}^2}$).  The most surprising finding is that despite having a series of magnetic interactions [Cr-$A$ and $A$-$A'$ with $A$, $A'$= Mn, Co, Ni, Cu], the $\theta_\mathrm{CW}$ of $A^5$Cr$_2$O$_4$ is comparable to  NiCr$_2$O$_4$~\cite{Suchomel:2012p054406} (also see Supplementary Figure S6) and MnCr$_2$O$_4$~\cite{Winkler:2009p104418}.  The effective magnetic moment also closely resembles  NiCr$_2$O$_4$ (6.64 $\mu_B$/f.u.). These findings also indicate that the net magnetic interaction, felt by a magnetic ion in CCO, can be treated using a mean-field approach.

To understand the effect of local distortions on magnetism, $|\theta_\mathrm{CW}|$ for all six compounds have been plotted as a function of EXAFS derived bond lengths and the $A$-site spin value in Fig.~\ref{fig: Fig 6}(b), (c). The $\theta_\mathrm{CW}$ of the parent compounds $A$Cr$_2$O$_4$ is obtained from Ref.~\cite{Lee:2000p3718,Suchomel:2012p054406, Winkler:2009p104418} and  an average of $A$-site spin ($S_A$=1.1) for $A^5$Cr$_2$O$_4$ has been considered for these plots. The  Cr-Cr bond length (Fig.\ref{fig: Fig 6} (b)) is considered as it corresponds to the Cr-Cr direct exchange. The Cr-O bond length is also plotted in connection with the $A$-O-Cr superexchange path (Fig.\ref{fig: Fig 6} (b)). While the $A$-O bond length also plays a role in  $A$-O-Cr superexchange strength, we were unable to create a similar plot for $A$-O bond lengths because this distance depends on the specific atom occupying the $A$-site in $A^5$Cr$_2$O$_4$. The $\theta_\mathrm{CW}$ of $A^5$Cr$_2$O$_4$ is closer to the value of NiCr$_2$O$_4$ and MnCr$_2$O$_4$. We have further evaluated the mean-field magnetic exchange parameters from our magnetization data, following the process described in Ref.~\cite{Winkler:2009p104418} [also see Supplementary section S4]. The exchange interaction parameters $J_{BB}$ (interactions between two Cr$^{3+}$), $J_{AB}$ (between $A^{2+}$ and Cr$^{3+}$),  and $J_{AA}$ (between two $A^{2+}$ cations) have been compared in Fig.~\ref{fig: Fig 6}(d). The exchange parameters for MnCr$_2$O$_4$ have been adapted from Ref.~\cite{Winkler:2009p104418}.  Interestingly, the $J_{BB}$ interaction shows almost no difference between these compounds. However, the  $J_{AB}$ and $J_{AA}$  values for $A^5$Cr$_2$O$_4$ are higher than those of MnCr$_2$O$_4$. The extraction of these $J$s using equation 2 is not possible for NiCr$_2$O$_4$ due to an anomaly in the magnetic susceptibility at the cubic-to-tetragonal transition ($\sim$ 315 K, see Supplementary Fig S6). However, Fig.\ref{fig: Fig 6} (b) and (c) clearly demonstrate the close similarity between $A^5$Cr$_2$O$_4$ and NiCr$_2$O$_4$ in terms of key parameters like Cr-Cr distance, Cr-O distance, and $A$-site spin and  $\theta_\mathrm{CW}$. This strong resemblance highlights the crucial role of local bond lengths around the Cr ions in determining the mean-field magnetic interactions within $A^5$Cr$_2$O$_4$.

Our DFT calculations also show that the local Cr moment increases with Cr-Cr and Cr-$A$ distances (see Supplementary Figure S7). This indicates that the moments become more localized as they are placed further away.
A more rigorous analysis of the connection between structure and magnetism may be obtained from the variation of the magnetic exchange parameters~\cite{Ganguly:2015p224417}. The extraction of pairwise exchange parameters for such a complicated disordered system is a highly non-trivial task and will be attempted to be done in the near future.

\section{Conclusions}
We have synthesized a compositionally complex spinel oxide $A^5$Cr$_2$O$_4$, with  Cr$^{3+}$ ions occupying the octahedral site and a uniform distribution of  $A^{2+}$ ($A$ = Mn, Co, Ni, Cu, and Zn) ions in the tetrahedral site.  Our investigation of the local structure employing element-specific EXAFS analysis and DFT calculations with various disordered configurations reveals several unique features that distinguish it from previously reported high entropy oxides. Unlike HEOs with rock-salt structures~\cite{Rost:2017p2732,Rak:2018p300}, where local Jahn-Teller distortions around Cu$^{2+}$ ions are typically preserved, our study reveals a complete absence of such distortions in $A^5$Cr$_2$O$_4$. In contrast to other HEOs, where local distortions are limited to the first nearest neighbors of the doped cations~\cite{Rost:2017p2732,Pu:2023peadi8809}, in the present study, $A^5$Cr$_2$O$_4$ exhibits variation in local distortions extending beyond the first neighbors. Despite a broad distribution of Cu-O bond distances, other $A$-O distances acquire specific values, implying the high flexibility of CuO$_4$ tetrahedral units to maintain an overall cubic structure by adjusting their positions. Such adjustment also leads to a broad distribution of second neighboring Cu-Cr distances and a narrow distribution of other $A$-Cr, Cr-Cr and Cr-O bond lengths. Despite substantial local distortions, the mean-field magnetic interaction energies of $A^5$Cr$_2$O$_4$ are remarkably similar to those of NiCr$_2$O$_4$ due to their closely matched average $A$-site spin value, as well as their average Cr-O and Cr-Cr bond lengths.

 Unveiling the temperature dependence of $A^5$Cr$_2$O$_4$'s long-range and local structures presents a compelling avenue for future research as some of the constituent members undergo structural transitions below room temperature~\cite{Suchomel:2012p054406}. Future investigations into the spin arrangements and potential multiferroic phases are warranted.  Furthermore, we believe that our  approach, which compares EXAFS analysis of CCO with the constituent parent members along with the DFT calculations for a set of disordered configurations, would be a powerful tool for revealing subtle details regarding structural modifications in other compositionally complex materials.

\subsection*{Methods}
{\bf Synthesis:} Polycrystalline samples of five parent compounds  $A$Cr$_2$O$_4$  (\textit{A} = Mn, Co, Ni, Cu and Zn)  were synthesized by conventional solid-state synthesis with a stoichiometric amount of Cr$_2$O$_3$ and \textit{A}O. The first heating was carried out at 900$^0$ C, followed by a second heat treatment at 1300$^0$ C with intermediate grindings. The annealing of  MnCr$_2$O$_4$ and CuCr$_2$O$_4$ were performed in argon and oxygen atmosphere, respectively, while the rest of the compounds were heated in air. The $A^5$Cr$_2$O$_4$ was synthesized in a similar route in the air.

{\bf Experimental details:} All the experiments were carried out at ambient temperatures. The sample purity of all samples was checked by powder XRD using a laboratory-based Rigaku Smartlab diffractometer. The final XRD patterns were further refined by the Rietveld method using the FULLPROF suite~\cite{CARVAJAL:1993p55}. The Raman spectra were recorded using Confocal Photoluminescence Raman Spectro Microscope. Data was collected using 532 nm LASER and 1800 rules/mm grating in 100-800 cm$^{-1}$ wavenumber range. The temperature-dependent magnetization measurement was carried out in the range of 5 K to 400 K using a commercial SQUID-VSM MPMS from M/s Quantum Design, USA.

XAS spectra for $L_{3,2}$ edges of Cr, Mn, Co, Ni, Cu of the $A^5$Cr$_2$O$_4$ were recorded in total electron yield mode at the beamline 4.0.2  of the Advanced Light Source, USA. Ni $L_{3,2}$ edge of NiCr$_2$O$_4$ and Cu $L_{3,2}$ edge of CuCr$_2$O$_4$ were also measured.

EXAFS measurements for all transition metal $K$-edge for all five parent compounds and $A^5$Cr$_2$O$_4$ have been performed at the P65 beamline, PETRA III Synchrotron Source (DESY, Hamburg, Germany) and at beamline 20-BM of the Advanced Photon Source, USA.
For these transmission mode experiments, the absorbers were prepared by uniformly coating fine powders on scotch tape. The incident and transmitted photon energies were recorded simultaneously using gas ionization chambers as detectors. For each \textit{K} edge, at least three scans were taken to average the statistical noise. All EXAFS data presented in this manuscript were acquired during the same beamtime at the P65 beamline. Therefore, the experimental setup was identical for all parent $A$Cr$_2$O$_4$ and $A^5$Cr$_2$O$_4$ samples.

{\bf EXAFS analysis details:} The analysis was conducted using well-established procedures from the DEMETER Suite\cite{Ravel:2005p537}. The spectra for each transition metal edge were calibrated with the corresponding standard metal foils. In pristine $A$Cr$_2$O$_4$ oxides, we performed fittings for the \textit{A} K-edge spectra (\textit{A} = Mn, Co, Ni, Cu, and Zn), using a \textit{k} range of 2 to 14 \AA$^{-1}$ and \textit{R} range of 1 to 4.5 \AA. For the Cr \textit{K}-edge, we employed a \textit{k} range of 2 to 12 \AA$^{-1}$ and an \textit{R} range of 1 to 3.7 \AA\ for the fittings. To prevent interference from the adjacent absorption edge in $A^5$Cr$_2$O$_4$, we truncated the spectra above 12 \AA$^{-1}$ in \textit{k}-space. The fittings were performed in the  \textit{k} range of 2 to 11 \AA$^{-1}$ and \textit{R} range of 1 to 3.7 \AA\ in all six edges (see Fig. 4(a), Fig. 5 (a), Fig. S3, and Tables S2-S4 in the Supplementary Materials).  In our fitting process, we primarily considered the following parameters: coordination number (\textit{N}), amplitude reduction factor ($S_0^2$), energy shift ($E_0$), the change in bond length ($\Delta R$), and the mean squared displacement ($\sigma^2$), also known as the Debye factor.


For the\textit{ A K} edge fitting in cubic \textit{A}Cr$_2$O$_4$ (\textit{A} = Mn, Co, Zn), we utilized four paths associated with the direct scattering: the first neighbor \textit{A}-O, the second neighbor \textit{A}-Cr, the third neighbor \textit{A}-O, and the fourth neighbor \textit{A}-\textit{A}. The parameters $S_0^2$  and $E_0$ were taken as variables for the fitting. $\Delta R$ and $\sigma^2$ were varied for each path considered. As a result, we had four values for $\Delta R$, four values for $\sigma^2$, one value for $S_0^2$, and one value for $E_0$, totaling ten parameters. For the Cr \textit{K} fitting, we focused on three coordination shells: Cr-O, Cr-Cr, and Cr-\textit{A}. Using a similar approach, we employed a total of eight parameters for this fitting. It is important to note that the \textit{A} \textit{K} and Cr \textit{K} fittings were conducted separately, and the parameters used in the two fittings are not correlated.

In CuCr$_2$O$_4$, at the Cu \textit{K} edge, the Jahn-Teller distortion leads to splitting in the second and third coordination shells of Cu-Cr and Cu-O respectively. Hence, two different values for $\Delta R$ and $\sigma^2$ were used within a single shell for both of these paths. As a result, six paths involving direct scatterers were used in the fitting, compared to four paths in the cubic phase. This change increased the total number of parameters from ten in the cubic phase to fourteen in the tetragonal phase. For the Cr \textit{K} edge, the Jahn-Teller distortion causes splitting in all three coordination shells: Cr-O, Cr-Cr, and Cr-Cu. Here, we used six paths in the fitting, instead of three in the cubic phase, which increased the number of total parameters from eight to fourteen. In the case of NiCr$_2$O$_4$, the Jahn-Teller splitting is smaller than the EXAFS resolution [$\Delta R$ = $\pi/2\times(k_{max} - k_{min})$], which is approximately 0.16 \textup{~\AA}. Since XRD indicated a tetragonal phase, we adopted a tetragonal model similar to that of CuCr$_2$O$_4$ to ensure consistency with the XRD results.

The coordination number \textit{N} was kept fixed during the analysis. All parent compounds are normal spinels, ruling out any site occupancy disorder between the \textit{A} (tetrahedral) and Cr (octahedral) site (second neighbors). However, we did try fittings by varying the coordination numbers of the first shell (\textit{A}-O and Cr-O). Although a small deviation from the stoichiometric value was observed, this deviation was consistent between all the parent compounds and the compositionally complex $A^5$Cr$_2$O$_4$ (see section S2.3 and Table S5 in Supplementary). Furthermore, the coordination numbers obtained from fitting agree with the stoichiometric values within the statistical error of the fit. Given the excellent agreement between the magnetic properties of our parent compounds and previously published results, we attribute these small deviations in oxygen coordination number to the inherent resolution limitations of EXAFS fitting.

In $A^5$Cr$_2$O$_4$, our XAS measurements and XANES spectra confirm a normal spinel structure. These confirm the presence of all \textit{A} (Mn, Co, Ni, Cu, and Zn) cations at the tetrahedral site, similar to all the parent compounds.   Therefore, Cr was modeled in octahedral site and the \textit{A} ions in tetrahedral site for the EXAFS fitting. For the fitting of the \textit{A}-site EXAFS in $A^5$Cr$_2$O$_4$, we chose the $S_0^2$ value determined from the fitting of the corresponding \textit{A}-site in the parent compound \textit{A}Cr$_2$O$_4$. In line with the spinel structure, we considered the following direct scattering paths for the \textit{A K} absorber: first neighbor pair \textit{A}-O (coordination 4), second neighbor \textit{A}-Cr (12), third neighbor again \textit{A}-O (12), and fourth neighbor \textit{A-A} (4). In this compound, there is a potential presence of five \textit{A} elements (Mn, Co, Ni, Cu, and Zn) in the fourth coordination shell with a total coordination number of 4. For instance, in the Co \textit{K} EXAFS, we have the fourth neighbor pairs as Co-Mn, Co-Ni, Co-Cu, Co-Zn, and Co-Co. Hence, in the fittings of every \textit{A} \textit{K} edge we used all the five pairs as fourth neighbors each with a coordination of 0.8.

In the Cr \textit{K} EXAFS fitting of $A^5$Cr$_2$O$_4$, we allowed $S_0^2$ to vary. We used the following direct scattering paths: the first neighbor pair Cr-O (6), the second Cr-Cr (6), and the third Cr-\textit{A} (6). In this case, the third neighbor, Cr-\textit{A}, has a coordination number of 6, and five \textit{A} elements can be present. Hence, we considered the pairs Cr-Mn, Cr-Co, Cr-Ni, Cr-Cu, and Cr-Zn, each of which with a coordination of 1.2.

In addition, we also checked with various combinations of elements and coordination at the \textit{A} site as the fourth neighbor from \textit{A} and third neighbor from Cr, finding no significant variations in the bond lengths, reported within the main text.

{\bf DFT calculations:} A cubic spinel structure of $A^5$Cr$_2$O$_4$ was utilized to generate ten different random oxide supercells of 2 $\times$ 2 $\times$ 2 dimensions, consisting of 448 atoms (64 $A$, 128 Cr, and 256 O atoms). To form the random distribution, 12 Mn, 13 Co, 13 Ni, 13 Cu, and 13 Zn atoms were distributed among the 64 available $A$ sites. Each of the ten random oxides was created by randomly rearranging the $A$ sites while keeping the number of each type of atom constant
We conducted structural optimization using the Vienna Ab initio Simulation Package (VASP) based on plane wave basis set and projector augmented wave pseudopotentials \cite{Kresse:1999p1758, Kresse:1996p11169}. The POSCAR files of the ten different structures have been supplied as a separate file along with the manuscript.
The exchange-correlation potential was described using the generalized gradient approximation (GGA) with the Perdew, Burke, and Ernzerhof (PBE) functional \cite{Perdew:1996p3865}. A static Hubbard correction was effectively applied to account for localized Coulomb interactions in highly correlated 3d transition metals, following the Dudarev approach \cite{Dudarev:1998p1505}. The $U_{eff}$ values were set to 4.0, 3.0, 5.0, 8.0, 8.0, and 3.0 eV for Mn, Co, Ni, Cu, Zn, and Cr, respectively, in line with the reported values for transition metal oxides \cite{Das:2015p425001, Wang:2006p195107, Himmetoglu:2011p115108, Harun:2020p102829}. This setup resulted in a good agreement in bond lengths between experimental data and DFT calculations. For Brillouin zone integration, a Gamma-centered 1 × 1 × 1 k-point grid was employed. The lattice constants were determined based on experimental results. The equilibrium atomic positions were obtained via energy minimization, employing the conjugate gradient method until the force components on each atom fell below 0.01 eV/Å.

\section*{Data availability}
	All data are available from the corresponding authors upon reasonable request,


\section*{Acknowledgements}
 The authors acknowledge the uses of central facilities of the department of Physics, IISc Bangalore, funded by the FIST program of DST (Department of Science and Technology), India. SM acknowledges funding support from a SERB Core Research grant (Grant No.
CRG/2022/001906). RN thanks the Indian Institute of
Science for support through Sir C. V. Raman postdoctoral fellowship program and Prof. K.R.S. Priolkar for commenting on the manuscript. SD and NB acknowledge funding from the Prime Minister’s Research Fellowship (PMRF), MoE, Government of India.
UGC-DAE Consortium for Scientific Research, Kalpakkam Node, Kokilamedu is acknowledged for providing SQUID-VSM MPMS facility for conducting magnetic measurements, which is part of this work.  Portions of this research were carried out at the light source PETRA III DESY, a member of the Helmholtz Association (HGF).  Financial support by the Department of Science \& Technology (Government of India) provided within the framework of the India@DESY collaboration is gratefully acknowledged.  This
research used resources of the Advanced Photon Source, a U.S. Department of Energy Office of Science User Facility operated by Argonne National Laboratory under Contract No. DE-AC02-06CH11357.  This research used resources of the Advanced Light Source, which is a Department of Energy Office of Science User Facility under Contract No. DE-AC02-05CH11231.
B.S. acknowledges financial support from Swedish Research Council (grant no. 2022-04309) and STINT Mobility Grant for Internationalization (grant no. MG2022-9386).The computations were enabled by resources provided by the National Academic Infrastructure for Supercomputing in Sweden (NAISS) at UPPMAX (NAISS 2024/5-258) and at NSC and PDC (NAISS 2023/3-42) partially funded by the Swedish Research Council through grant agreement no. 2022-06725. B.S. also acknowledges the allocation of supercomputing hours granted by the EuroHPC JU Development Access call in LUMI-C supercomputer (grant no. EHPC-DEV-2024D04-071) in Finland. S.E. acknowledges the allocation of supercomputing hours granted by the EuroHPC JU Development Access call in LUMI-C supercomputer (grant no. EHPC-DEV-2024D03-043) in Finland and Karolina supercomputer (grant no. EU2023D11-039) in Czech Republic.

\section*{Author contribution}

SM conceived and supervised the project. RN and SD synthesized, characterized the samples and analyzed all results. RN, SD, NB, CK carried out soft X-ray absorption spectroscopy measurements. RN, NB, TD, GES performed EXAFS experiments. S.E. has performed all the calculations based on DFT, and B.S. has supervised the theoretical part. RN and SM wrote the manuscript. All authors discussed the results and commented on the manuscript.

\section*{Competing interests}

	The authors declare no competing interests.

\end{document}